# Self-consistent spin-fluctuation spectrum and correlated electronic structure of actinides

Tanmoy Das, Jian-Xin Zhu, Matthias J. Graf

Theoretical Division, Los Alamos National Laboratory, Los Alamos, NM 87545, USA

(Dated: September 5, 2012)

We present an overview of various theoretical methods with detailed emphasis on an intermediate Coulomb-$U$ coupling model. This model is based on material-specific *ab-initio* band structure from which correlation effects are computed via self-consistent $GW$-based self-energy corrections arising from spin-fluctuations. We apply this approach to four isostructural intermetallic actinides $PuCoIn_5$, $PuCoGa_5$, $PuRhGa_5$ belonging to the Pu-115 family, and $UCoGa_5$, a member of the U-115 family. The 115 families share the property of spin-orbit split density of states enabling substantial spin fluctuations around 0.5 eV, whose feedback effect on the electronic structure create mass renormalization and electronic 'hot spots', i.e., regions of large spectral weight. A detailed comparison is provided for the angle-resolved and angle-integrated photoemission spectra and de Haas-van Alphen experimental data as available. The results suggest that this class of actinides is adequately described by the intermediate Coulomb interaction regime, where both itinerant and incoherent features coexist in the electronic structure.

# I. INTRODUCTION

Dynamic correlation effects of strongly correlated electron systems pose a serious challenge to our theoretical understanding of the physical behavior of these materials. Strong renormalization of electronic bands and their spectral weight anomaly cannot be accounted for by the density functional theory. Along this line, rare-earth 4$f$ and actinide 5$f$ electron systems are particularly challenging due to their low-energy fermionic excitations with heavy electron mass. A popular model that accounts for this phenomenon is the Anderson model,[1] where the interaction between localized $f$ electron spins and itinerant conduction electrons (mainly stemming from $d$-orbitals) leads to heavy mass and an interplay between Kondo physics, magnetism and superconductivity.[2] The resulting interaction pushes the system to the intermediate Coulomb-$U$ region in which neither the purely itinerant mean-field theory nor the strong-coupling Anderson lattice model holds exactly. This twilight zone of intermediate coupling gives rise to prototypical examples of strongly correlated electron systems,[3] and provides a platform for the emergence and coexistence of multiple competing phases of matter. The observation of universal scaling of the spin-fluctuation temperature $T_s$ versus superconducting transition temperature $T_c$ supports further this picture that the Pu-115 compounds lie in between the Ce-based 4$f$-electron heavy-fermion and $d$-electron superconductors (cuprates and pnictides).[4]

An advantage of the intermediate coupling method is that the diagrammatic perturbation theory of fluctuations can still be applied as long as the Hubbard (Coulomb) $U \sim D$, where $D$ is the non-interacting bandwidth.[5,6] Here one starts from the itinerant band picture of electrons where dynamic correlation effects are added perturbatively, giving rise to strong electronic structure renormalization. The fluctuation theory becomes exact in the weak coupling regime, but is rather complicated to treat in the intermediate regime. Various approaches have been designed to deal with this regime.[7] In this feature article, we present an overview of the intermediate Coulomb-$U$ coupling model based on the self-consistent $GW$ formalism of the self-energy within random phase approximation (RPA), and the obtained results will be compared directly with various spectroscopies. The material specific *ab-initio* band structure is used as input to the calculations, which plays a very important role in deriving quantitative agreement with experiments. Not only does it enable us to obtain a systematic description of the correlation spectrum, but also helps us to study how the high-energy correlation spectrum evolves naturally into magnetic, and superconducting ground states, which vary dramatically within the same 115 actinide family.[8–10]

Uranium-based 115 compounds are often contrasted with the rich phase diagram of the isostructural Pu-115 compounds: PuCoGa$_5$,[11] PuRhGa$_5$,[12] and PuCoIn$_5$ [Ref. 13] were discovered to be superconducting with characteristically varying transition temperature $T_c$, whereas U$T$Ga$_5$ ($T$: Fe, Co, Pt)[14,15] are magnetic and UCoGa$_5$ is a paramagnet,[6] yet not a typical Fermi liquid.[16] Due to this diversity of ground states within the same family, it provides an excellent test bed for applying our approach. In future design of materials properties one might even be able to correlate specific electronic 'hot spots' in the spectral function with specific ground state properties for controlled functionality.

From experimental side, the photoemission data reveal a strong spectral weight redistribution in the single-particle spectrum with a prominent peak-dip-hump structure around 0.5 eV in Pu-115[17] and similarly in UCoGa$_5$, UNiGa$_5$ and UFeGa$_5$.[18] This feature was earlier assigned within a phenomenological picture of $f$-electron duality as the separation between itinerant (peak) and incoherent (hump) states of the 5$f$ electrons.[19,20] We interpret these spectra in terms of the spin-fluctuation properties creating a dip in the dressed quasiparticle excitations due to strong scattering in the particle-hole continuum. The lost spectral weight (dip) is partially distributed to the renormalized itinerant states at the Fermi level (peak), as well as to the strongly localized incoherent states at higher energy (hump). The coherent states at the Fermi level can still be characterized as Bloch waves, though renormalized, whereas the incoherent electrons appear localized in real space exhibiting the dispersionless hump feature.

The rest of this feature article is arranged in the following way. We start with a recapitulation of the existing models for actinides in Sec. IIA, and then we summarize our intermediate coupling model in Sec. IIB. In Sec. III, a*b-initio* band structure, without correlation, of Pu-115 and UCoGa$_5$ family is discussed, followed by results for the momentum-resolved, and momentum integrated spin-fluctuation spectrum, self-energy functional, and the corresponding dressed electronic structure. The calculations are compared with integrated (PES) and angular-resolved photoemission spectroscopy (ARPES), as well as with de Haas-van Alphen (dHvA) measurements as available in Sec. IV. A ubiquitous nature of these behaviors between actinide, high-temperature cuprate superconductors, and strontium-ruthenates is also deduced in Sec. IVC. In Sec. V, we discuss the implication of this model, and results to the low-energy mass-renormalization and unconventional superconductivity. Conclusions and related future directions are given in Sec. VI. Finally, we relate the calculated spin-fluctuation coupling constant λ to the superconducting transition temperature ($T_c$) as we move across the family, suggesting that spin fluctuations also play a crucial role in the pairing mechanism.

## II. THEORETICAL METHODS

### A. Various models of correlated electron systems

With the rapidly growing computer technology, the development of density functional theory (DFT) [for a review see Ref. 21] has led to significant progress in electronic structure theory during the past five decades. In principle, the theory is rigorous in describing the ground state properties of electronic systems.[22] In practice, the exchange-correlation functional, which is the most crucial component of the theory, is designed with approximations. Among them, the local density approximation (LDA)[23] and generalized gradient approximation (GGA)[24] have turned out to be quite successful in describing "good" metals and a few semiconductors. However, the LDA-based DFT theory fails badly for two important classes of materials, partially filled $d$-electron based transition metal oxides and $f$-electron based lanthanides and actinides. These classes of materials are broadly termed as strongly correlated electronic systems, in which Landau's Fermi liquid description of dressed single particles often fails to characterize low-lying excitations. It is well known that the implemented correlation exchange functionals in the DFT underestimate strong correlation effects of electrons.

During the past two decades, significant progress has been made to overcome the inadequacy of the LDA and GGA methods for strongly correlated electron materials. A few correlated band approaches like LDA+$U$,[25] hybrid-density functional,[26] and self-interaction corrected local spin-density approximation have been proposed within the framework of DFT.[27–29] To truly capture the dynamical effects, many-body techniques, which were mostly developed in the many-body model Hamiltonian community, have been developed to include materials specifics. Two of the most notable and currently popular many-body approaches are the first-principle $GW$ approximation,[30,31] and the combination of the dynamical mean-field theory (DMFT)[32–34] with the LDA.[35] Within the $GW$ method, the self-energy can be formally expanded in powers of the screened interaction $W$, as such starting from the itinerant limit. Originally limited to the electron gas, many applications to real materials like semiconductors and insulators have been reported.[36] While within the LDA+DMFT method, the effects due to local Coulomb interaction $U$ are treated non-perturbatively by mapping a quantum many-body lattice problem onto an effective impurity problem with the conduction electron bath determined self-consistently.[34] In the single-site DMFT implementation, the self-energy on the correlated orbitals are assumed to be local, which captures quantum temporal fluctuations but not spatial fluctuations. In this sense, the LDA+DMFT method[35,37] is ideal for the description of systems sitting on the localization-delocalization border. The LDA+DMFT technique has been applied with remarkable success to transition metals, transition metal oxides, and some *f*-electron materials. For references on this vast topic see Refs. 35,37. More recently, various groups attacked the correlation problem of *f*-electrons in URu$_2$Si$_2$,[38] CeIrIn$_5$,[39,40] and in the Pu-115 family.[19,20,41,42] However, most correlated electron systems of interest are in the intermediate coupling regime, where a classification in neither weak nor strong regime applies. Which of these methods ultimately gives a better description of dynamic correlation effects is a priori unknown, and subject of ongoing research.

This review focuses on the intermediate Coulomb-*U* coupling model combined with electronic structure calculations of strongly correlated electron materials. We present the basic concepts and computational tools of the RPA based *GW* method for calculating spin-fluctuation dressed electronic states in the intermediate coupling regime, and apply it to the intermetallic actinide families of Pu-115 and U-115. We anticipate that with the continuing development of many-body methods, the increasing computational power, and development of novel algorithms, researchers will get new tools for material modeling that can treat effectively the band renormalization and low-excitation spectrum of strongly correlated electron materials as expressed in many actinides.

### B. An intermediate Coulomb-*U* coupling model

In the intermediate Coulomb-*U* coupling model we calculate material-specific properties affected by dynamic electron correlations. The input for the many-body self-energy calculations of electron correlations is based on first-principles electronic band structures, including spin-orbit coupling within the framework of density functional theory using the GGA.[43] The spectral representation of the Green's function is constructed as

$$G^0_{sp}(\boldsymbol{k}, i\omega_n) = \sum_\mu \frac{\varphi^s_\mu(\boldsymbol{k})\varphi^{p*}_\mu(\boldsymbol{k})}{i\omega_n - E_\mu(\boldsymbol{k})} \; . \tag{1.1}$$

Here $\varphi^i_\mu$ is the eigenstate for $\mu^{th}$-band $E_\mu$ projected on the $i^{th}$ orbital. The non-interacting susceptibility in the particle-hole channel is calculated by convoluting the Green's function over the entire Brillouin zone (spin and charge bare susceptibility are the same in the paramagnetic ground state):[44,45]

$$\chi^0_{spqr}(\boldsymbol{q},\Omega) = -\frac{T}{N}\sum_{\mathbf{k},n} G^0_{sp}(\boldsymbol{k},i\omega_n)G^0_{qr}(\boldsymbol{k}+\boldsymbol{q},i\omega_n+\Omega). \quad (1.2)$$

Within the RPA, spin and charge channels become decoupled (we ignore particle-particle as well as weaker charge fluctuation processes). In the spin channel, the collective many-body corrections of the spin-fluctuation spectrum in RPA can be written in matrix representation: $\hat{\chi} = \hat{\chi}^0[\hat{1} - \widehat{U}_s\hat{\chi}^0]^{-1}$. The interaction matrix $\widehat{U}_s$ is defined in the same basis consisting of intra-orbital $U$, inter-orbital $V$, Hund's coupling $J$ and pair scattering $J'$ terms.[45–47] In our definition, the interaction vertex is a tensor of dimension $n^2 \times n^2$ ($n$ is the number of bands) with $U$ and $V$ going to the diagonal terms, while $J$ and $J'$ going to the off diagonals. In the present calculation, we neglect the orbital overlap of eigenstates, i.e., we assume $\varphi^i_\mu$ when $i=\mu$. Such an approximation simplifies the calculation and $\hat{\chi}^0$ becomes a diagonal matrix with $J = J' = 0$.

For calculation of the many-body self-energy in the *GW* approximation, *G* represents the Green's function and *W* the interaction vertex. In this work, we are using a modified or '*GW*-like' version of the conventional *GW* approximation. First, we are performing only a 'single-shot' self-energy calculation, i.e., we are not feeding back the renormalized Green's function into the electronic structure calculation. Second, we are using a phenomenological on-site Coulomb $U$ in the spirit of the Anderson model of localized moments coupled to conduction electrons, instead of a self-consistent Coulomb potential. In this *GW*-like approximation we denote the vertex by *V*, instead of *W*, and write the matrix element of the spin-fluctuation interaction vertex in the fluctuation-exchange approximation following Ref. 48:

$$V_{pqrs}(\boldsymbol{q},\Omega) = \left[\frac{3}{2}\widehat{U}_s\hat{\chi}''(\boldsymbol{q},\Omega)\widehat{U}_s\right]_{pqrs} \quad (1.3)$$

The Feynmann-Dyson equation for the imaginary part of the self-energy in a multiband system with *N* sites is

$$\Sigma''_{pq}(\boldsymbol{k},\omega) = -\frac{1}{N}\sum_{rs}\sum_{\boldsymbol{q}}\int_{-\infty}^{\infty}d\Omega\,\Gamma(\boldsymbol{k},\boldsymbol{q},\omega,\Omega)\widehat{V}_{pqrs}(\boldsymbol{q},\Omega)\begin{bmatrix}(n_B(\Omega)+f(\omega+\Omega))A_{rs}(\boldsymbol{k}+\boldsymbol{q},\omega+\Omega)\\+(n_B(\Omega)+1-f(\omega-\Omega))A_{rs}(\boldsymbol{k}+\boldsymbol{q},\omega-\Omega)\end{bmatrix}, \quad (1.4)$$

The quasiparticle spectral function is defined by $A_{rs}(\boldsymbol{k},E) = -\text{Im}\,G^0_{rs}(\boldsymbol{k},E)/\pi$. The quantities $n_B$ and $f$ are the Bose-Einstein and Fermi-Dirac distribution functions, respectively. The vertex correction $\Gamma$ will be assumed isotropic as discussed later. For a more accurate calculation, one needs to account for the anisotropy in $\widehat{V}(\mathbf{q},\Omega)$. In the present case, where the spin-fluctuation spectrum is only weakly anisotropic (see Fig. 3), it is justified to replace the first term in Eq. (1.4) by a momentum-averaged spin-fluctuation function interaction, that is, $\langle \hat{V}(\Omega) \rangle_q = \left[\frac{a^3}{(2\pi)^3}\right] \int d^3q \hat{V}(q,\Omega)$. This is equivalent to dropping the $k$ dependence of the self-energy, which greatly simplifies the numerical self-consistency loop. It then follows from Eq. (1.4) that at $T=0$, the imaginary part of the self-energy reduces to

$$\Sigma''_{pq}(\omega) \approx -2\sum_{rs}\int_0^\omega d\Omega \langle V_{qprs}(\Omega) \rangle_q N_{rs}(\omega - \Omega), \qquad (1.5)$$

for $\omega > 0$, where the density of states is given by $N_{rs}(E) = \sum_k A_{rs}(k,E)$. (For $\omega < 0$, the only changes are that the upper limit of the integral is $|\omega|$ and the argument of $N_{rs}$ is $\Omega-|\omega|$, which is $< 0$.) We use Eq. (5) to compute the imaginary part of the self-energy from the first-principles band structure. The real part of the self-energy, $\Sigma'_{pq}(\omega)$, is obtained by using the Kramers-Kronig relationship. Finally, the self-energy dressed quasiparticle spectrum is determined by Dyson's equation:

$$\hat{G}^{-1}(k,\omega) = \hat{G}_0^{-1}(k,\omega) - \langle \hat{\Sigma}(k,\omega) \rangle_k. \qquad (1.6)$$

Full self-consistency requires the bare Green's function $\hat{G}_0$ to be replaced with the self-energy dressed $\hat{G}$ in Eq. (1.6) in $\hat{\chi}^0$ in Eq. (2) and in the spectral function $A$ in Eq. (1.5). This procedure is numerically expensive, especially in multiband systems. Therefore, following the proposal of Markiewicz *et al.*,[49] we adopt a modified self-consistency scheme, where we expand the real part of the self-energy $\Sigma'(\omega) = (1-Z^{-1})\omega$ in the low-energy region [$\omega < 200$ meV $\ll U$], where the imaginary part of the self-energy $\Sigma'' \sim 0$. The resulting self-energy dressed quasiparticle dispersions $\bar{E}_n(k) = ZE_n(k)$ are used in Eqs. (1.1)-(1.6), which keeps all the formalism unchanged with respect to the quasiparticle renormalization factor $Z$. Finally, in lowest order $Z$ introduces a vertex correction, which simplifies the complex vertex function to $\Gamma(k, q, \omega, \Omega) \approx Z^{-1}$ according to the Ward identity, relating the vertex function with the self-energy.[50]

Our choice of the screened Coulomb term $U$ satisfies the intermediate coupling approximation of $U/D \sim 1$. As seen from the band structures in Fig. 1(a), the average bare bandwidth for all materials near the Fermi level is of order of 1 eV. Hence, we set $U=1$ eV for all compounds, which is below the critical value of a magnetic instability, that is, $U\chi^0(q,\omega = 0) < 1$ for all $q$. Note that our screened $U$ for the spin-fluctuation calculation is smaller than that used in LDA+U type calculations, where a larger value of $U \sim 3$ eV was introduced into the local orbital basis.[41,42,51]

## III. RESULTS

### A. *Ab-initio* electronic structure

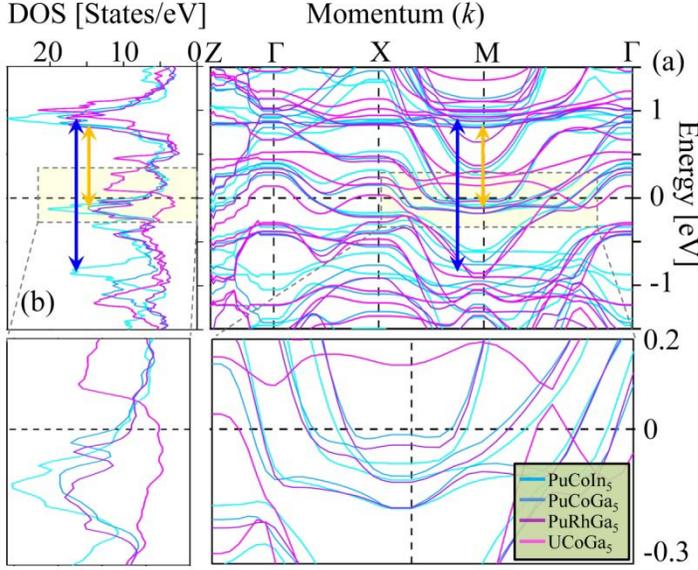

FIG. 1. (a) First-principles GGA electronic band-structure calculations for various Pu-115 and UCoGa$_5$ actinides near $E_F$.[6] (b) Corresponding DOS in the low-energy region of present interest. The double-arrows mark the relevant particle-hole excitations. Insets: Low-energy regions of dispersion and DOS showing that all materials are related by a rigid shift of bands in this energy scale.

First-principles calculations are carried out by using the full-potential linearized augmented plane wave (FP-LAPW) method as implemented in the WIEN2k code.[52] The GGA implementation[24] was used for exchange-correlation functionals. The spin-orbit coupling was included in a second variational way, for which relativistic $p_{1/2}$ local orbitals were added into the basis set for the description of the 6p states of uranium or plutonium.[53] For example, the energy spread to separate the localized valence states in UCoGa$_5$ was -6 Ryd. The muffin-tin radii were: $2.5 a_0$ for U, $2.48 a_0$ for Co, and $2.2 a_0$ for Ga, where $a_0$ is the Bohr radius. The criterion for the number of plane waves was $R_{MT}^{min} K^{max} = 8$ and the number of *k*-points was 40×40×25. The self-energy calculation below was performed with 40 bands to capture the ±10 eV energy window of relevance around the Fermi level.

Figure 1 presents the calculated GGA band structure in panel (a) and the corresponding non-interacting density of states (DOS) in (b) for all four materials studied here. Zooming into the Fermi level region in the *inset*, we notice that the low-energy band structure remains very much the same shape for all materials. It primarily shifts upward in energy when moving along the series PuCoIn$_5$→PuCoGa$_5$→PuRhGa$_5$→UCoGa$_5$. To leading order, this behavior can very much be accounted for by a rigid band shift. The Pu-115 compounds show two sharp peaks in the DOS just below and above the Fermi energy ($E_F$), which mainly originate from the 5f electrons of Pu atoms. This result is consistent with earlier calculations of the partial DOS in Refs. 20, 54, in which we find that the 3d (or 4d) and 4p (or 5p) electrons of the reservoir elements are not important in this energy scale. As the DOS at $E_F$ decreases in going to UCoGa$_5$ (see magenta lines in Fig. 1), most of the 5$f$ states move above $E_F$, thus reducing the correlation strength to a large extent.

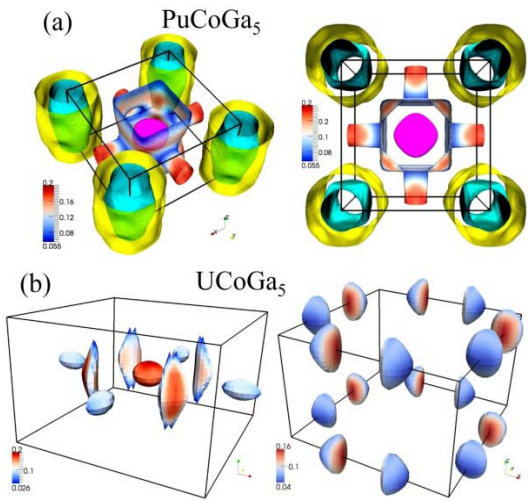

FIG. 2. The GGA calculated FSs of PuCoGa$_5$ and UCoGa$_5$. The top panel (a) gives the same four FSs for PuCoGa$_5$, plotted in angle-view (left) and top view (right). The FSs for two different bands of UCoGa$_5$ are separated, for clarity, into two figures in the bottom panel (b). The color coding (blue to red) reflects the Fermi speed (low to high) or the inverse of the density of states. The BZ boxes are centered at the Γ point.

The Fermi surface (FS) topologies for two representative systems, namely of PuCoGa$_5$ and UCoGa$_5$, are shown in Fig. 2. Despite the above mentioned overall similarities in the electronic dispersion the fermiology between these materials is significantly different as one might expect in a material-specific study. The fermiology of the other Pu-115s[20] are very much similar to PuCoGa$_5$ and thus are not shown. As demonstrated for the band structure and DOS in Fig. 1, the FS for Pu-115 systems consists of two cylinder-shaped pieces at the corners of the BZ with lesser three dimensionality. The FS of UCoGa$_5$ is dramatically smaller and highly dispersive along $k_z$. In fact for the latter case, none of the FS pieces survive at all values of $k_z$. These differences in FSs will be discussed in more details in Sec. VA. We argue that it is responsible for the absence of superconductivity in UCoGa$_5$. Finally the results will be compared with de Haas-van Alphen (dHvA) oscillations in Sec. IVE. We note that our GGA calculations agree with previous electronic structure calculations,[54,55] except for the detailed shape of the tube-like or pillow-like FS pockets of UCoGa$_5$, which are very sensitive to the location of $E_F$ within ∼ 13 meV (or ∼ 1 mRyd).

## B. Momentum and energy dependence of correlation function

An advantage of the present framework is that, in principle, it can incorporate the momentum dependence of the correlation vertex and vertex correction function. We present the full momentum dependence of the computed spin-fluctuation vertex, $\hat{V}(\boldsymbol{q},\omega)$, in Fig. 3 plotted as a function of excitation energy along the high-symmetry momentum directions. We see that in some cases the momentum-dependence can be large, and the maximum intensity can shift along the $q_z$ direction. Corresponding momentum-averaged values $<V>_q$ are fairly similar for all Pu-115 compounds, but notably different for UCoGa$_5$. The low-energy peak arises from transitions between the 5$f$ states lying just below and above $E_F$ (within the RPA, the peak shifts to lower-energy), see gold arrow in Figs. 1(b) and 3(e). The high-energy hump comes mostly from the transition of the second peak in the DOS below $E_F$ (hybridized *d-* and *p-*states also contribute[54]) to the 5*f* states above $E_F$ as marked by the blue arrow in Figs. 1(b) and 3(e). For UCoGa$_5$ most of the 5*f* states shift above $E_F$ and thus intra-orbital spin fluctuations do not survive, while the inter-orbital spin fluctuations move to higher energy.

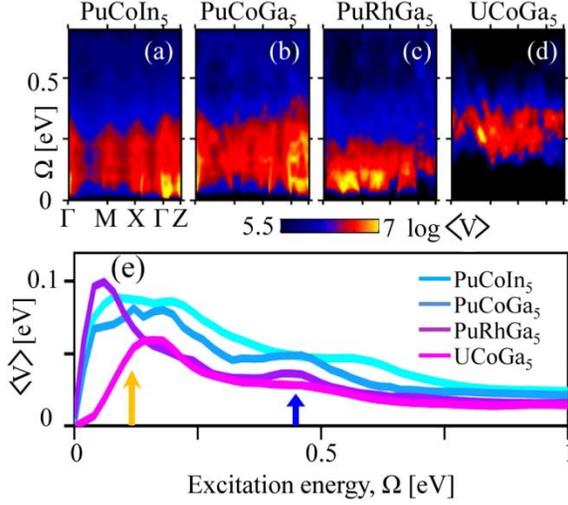

FIG. 3. The momentum- and energy-dependence of the spin-fluctuation vertex $V(\boldsymbol{q},\Omega)$ is plotted along high-symmetry directions in (a)-(d). Panel (e): The corresponding $\langle V(\Omega)\rangle_q$ averaged over 3D momentum space. All calculations are performed for ±10 eV energy window, but results are shown only in the relevant energy region. Reproduced from Ref. 6. Arrows dictate two peaks in the *f-f* (low-energy) and *f-d* (higher energy) transition channels.

### C. Energy-dependent self-energy and band renormalization

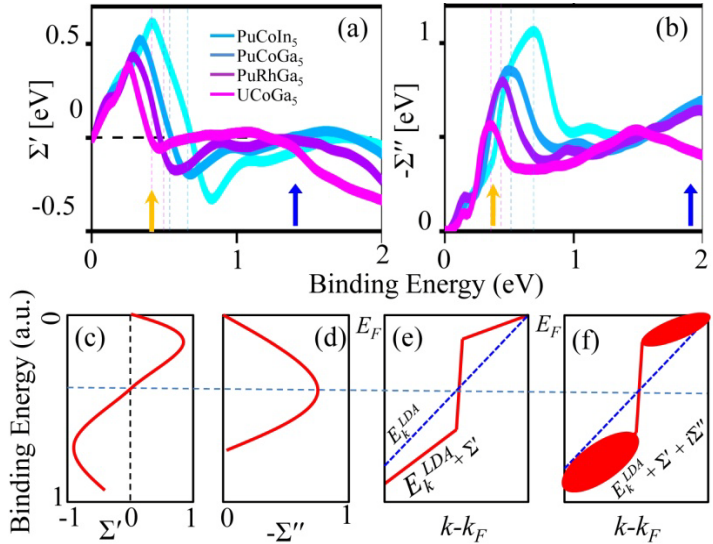

FIG. 4. The computed momentum-averaged $\Sigma'(\omega)$ and $\Sigma''(\omega)$ are plotted in (a) and (b), respectively. All peak positions in $\langle V\rangle$ in Fig. 3(b) are shifted to higher energy in $\Sigma''$ in (b) due to band-structure effects. (c)-(d) Schematic energy dependence of the general shape of the real and imaginary part of the self-energy. $\Sigma'$ contains a sign-reversal at a characteristic energy $\Sigma''$ is expected to obtain a peak at the energy. (e) A linear dispersion is shown to be renormalized by $\Sigma'$. (f) Typical form of the spectral weight map in the energy and momentum space due to full $\Sigma$ -correction. Arrows have same meanings as Fig. 3

The coupling of the spin fluctuations to the electron dynamics gives the self-energy correction in Eq. (4). The real and imaginary parts of the self-energy $\Sigma(\omega) = \langle \Sigma(\boldsymbol{k},\omega)\rangle_k$ are plotted in Figs. 4(a) and 4(b), respectively. Note that $\Sigma''(\omega)$ shows a peak-dip-hump feature, which is strongly enhanced by the electronic state in comparison with $\langle V(\boldsymbol{q},\Omega)\rangle_q$. Both the low-and high-energy features move toward $\omega = 0$ as the 5*f* states shift toward $E_F$ across the series PuCoIn$_5$→PuCoGa$_5$→PuRhGa$_5$→UCoGa$_5$ (for UCoGa$_5$ the 5*f* states eventually move above $E_F$). The itinerant and localized picture of the quasiparticle character in this class of materials is intimately related to the shape of the self-energies, as schematically illustrated in the lower panel of Fig. 4. In these systems, the strength of the $\Sigma''(\omega)$ peak acquires values that split the overall quasiparticle states into two energy scales; at low energies it induces itinerant quasiparticle states, while at higher energies it creates incoherent localized states. The resulting $\Sigma'(\omega)$ acquires an interesting energy dependence. The peak height of $\Sigma''(\omega)$ is sufficiently large such that it leads to $\Sigma'(\omega) = 0$ at this energy, which is possible within the intermediate Coulomb interaction scenario, but not in the weak-coupling regime. Below this crossover energy $\Sigma'(\omega) > 0$, which causes the quasiparticles to acquire mass enhancement with quasiparticle weight Z< 1. As a consequence the corresponding bands are renormalized toward the Fermi level. Above this characteristic energy $\Sigma''(\omega)$ causes the overall quasiparticle state to sharply drop from Z< 1 to Z> 1 region. At the crossover energy, the spectral weight is strongly suppressed due to the peak in $\Sigma''(\omega)$. The obtained shape of the quasiparticle dispersion is dubbed 'waterfall' or 'S' feature due to both its visual and conceptual similarity;[49,56–59] see also Sec. IV C.

### D. Momentum-resolved single-particle spectrum

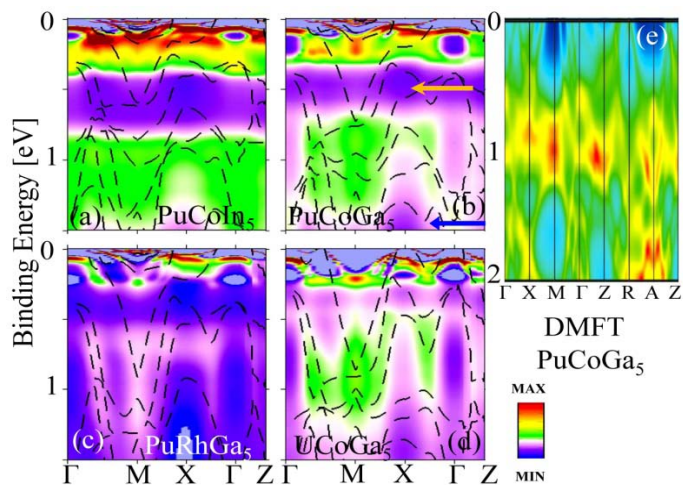

FIG. 5. (a)-(d): The self-energy dressed angle-resolved spectral weight $A(\mathbf{k},\omega)$ is plotted along high-symmetry momentum directions. The peak-dip-hump feature is clearly evident in all spectra below 1 eV. Reproduced from Ref. 6. The LDA+DMFT calculation of $PuCoGa_5$ from Ref. 41 is shown in (e) for comparison. We see that DMFT predicts the pile-up of the spectral weight around 1 eV (lower Hubbard band) with virtually no spectral weight at lower energy. The experimental data in Fig. 7 and in Fig. 10, however, demonstrate substantial spectral weight in the low-energy region, which agrees well with the present calculation. Two arrows in panel (b) indicate generic dips in the spectral weight, related to peaks in the self-energy.

The self-energy dressed quasiparticle state is shown in Fig. 5 for all four compounds as a function of energy and momentum. In all spectra the pile-up of spectral weight (hot spots) at two or three energy scales is clearly visible. A more detailed momentum resolution of the data will be given below. At low energies, when $\Sigma' > 0$, all quasiparticle states are renormalized toward $E_F$. In this energy region $\Sigma''$ is small, reflecting that quasiparticle states are coherent and itinerant. Above the peak in $\Sigma''$, where $\Sigma' < 0$, quasiparticle states are pushed to higher energy. The lost spectral weight from the peak in $\Sigma''$ is redistributed in both lower and higher binding energies. A similar spectral weight redistribution occurs at the second peak (hump) in $\Sigma''$ near 2 eV binding energy. As a result further pile-up of spectral weight occurs around 1.0-1.5 eV, creating hot spots of new quasiparticle states due to electronic correlations. The quasiparticle states in this energy region are incoherent and fairly dispersionless, reflecting the dual aspect of the localized behavior of 5f electrons. Qualitatively similar behavior was also found by using the LDA+DMFT method[41] shown in Fig. 5(e), however, with a weaker renormalization, and significantly less spectral weight near $E_F$ compared to experiments (Figs. 7 and 10).

### E. Peak-dip-hump feature

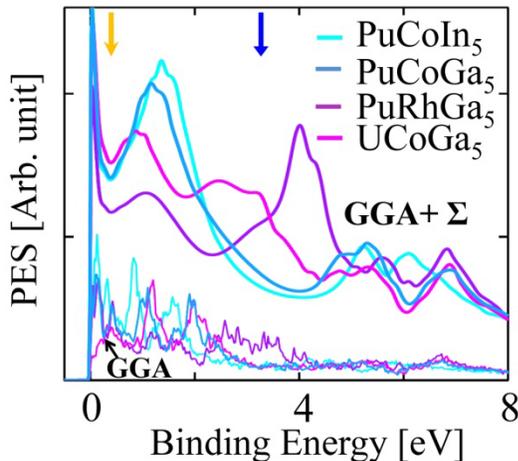

FIG. 6. (a) Computed PES spectra for various Pu-115 and U-115 compounds. All theoretical spectra have been renormalized by the same scaling factor. The GGA DOS is plotted with a much reduced normalization (taken from Fig. 1(b)) for comparison. Two arrows indicate generic dips in the spectral weight, related to peaks in the self-energy in Fig. 4.

The duality or coexistence of itinerant and localized quasiparticles manifests itself as a prominent peak-dip-hump feature in the momentum-integrated spectral function or DOS: $I_{PES} = \langle A(\mathbf{k},\omega)\rangle n_F(\omega)$, where we neglect for simplicity any matrix-element effects. The results are shown in Fig. 6. As we move across the series from PuCoIn$_5$ to UCoGa$_5$ the spectral weight redistribution gradually decreases. This suggests that spin fluctuations play a lesser role in UCoGa$_5$ than in the isostructural Pu-115 compounds. We show a direct comparison of our theory with available experimental data in Fig. 10 below.

### IV. COMPARISON WITH EXPERIMENTS

### A. Angle-resolved photoemission spectroscopy

The subsequent ARPES measurement[16] in UCoGa$_5$ revealed the anomalous momentum and energy dependence of the electronic dispersions shown in Fig. 7, which is in close agreement with the key features of our predictions. Figs. 7(a) and 7(b) show the 3D and 2D rendering of the same data, respectively. The dispersion (traces of the intensity peaks), in

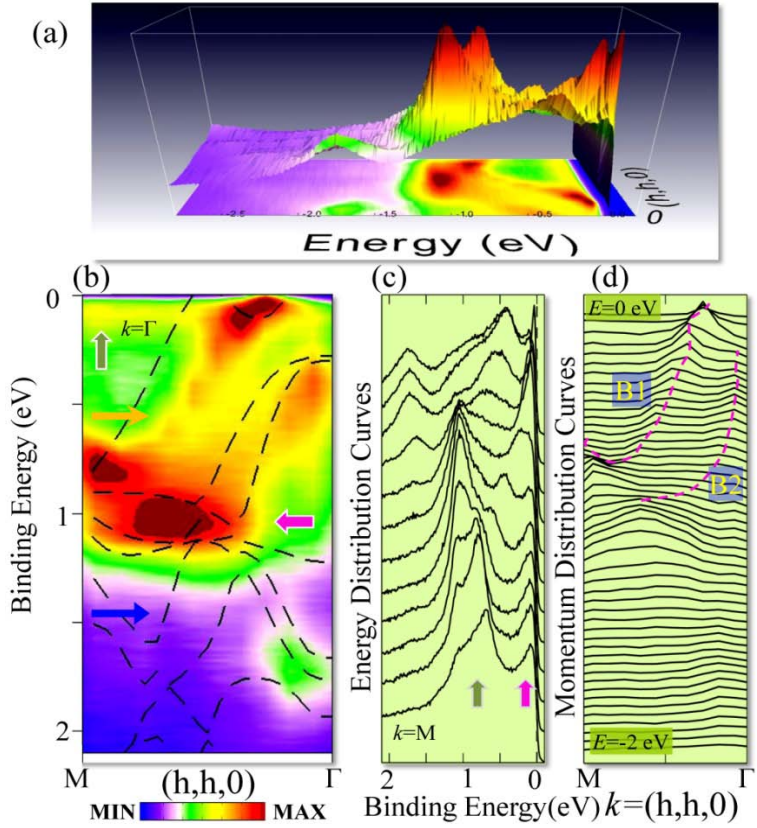

FIG. 7. Dispersion and measured ARPES spectral function anomaly. (a) 3D intensity map of UCoGa$_5$ along the M→Γ direction ($h$, $h$, 0) in the Brillouin zone. (b) Same data plotted as 2D contour map and compared to the corresponding *ab-initio* GGA electronic band-structure dispersions (black dashed lines). Arrows indicate characteristic features in the dispersion and spectral weight such as regions of dip or peak (see text). All the ARPES data are collected at the photon energy of 46 eV, with energy resolution better than 20 meV. (c) The EDCs are plotted for fixed momenta. The curves from bottom to top are chosen for equally spaced momentum points from M to Γ. Arrows have same meaning as in (a). (d) MDCs for fixed binding energy. Bottom to top curves are chosen from $E = -2.1$ eV to $E = 0$ eV. The dashed lines are guide to the eyes for two extracted low-lying dispersions of anomalous character, analyzed in Fig. 2.

Fig. 7(b), reveal a drastic departure of the quasiparticle states from the *ab-initio* electronic structure calculations (dashed lines). More importantly, the associated quasiparticle width at the peak positions is significantly momentum and energy dependent. Unlike the renormalized quasiparticle dispersion relation, which is a genuine manifestation of many-body correlation effects, the anomaly in the lifetime may have many intrinsic and extrinsic causes and is more difficult to assign to a particular mechanism.[60] We notice that while bare GGA bands are present in the entire energy and momentum region, the spectral intensity is accumulated mainly at two energy scales. As mentioned earlier, this connection between low and high energy scales is analogous to the so-called 'waterfall' or high-energy-kink feature observed in single-band cuprates[56–58] and Sr$_2$RuO$_4$[61] as compared in Fig. 9, see below. The energy-distribution curves (EDCs) of the ARPES intensity at several representative fixed momenta in Fig. 7(c), and momentum-distribution curves (MDCs) at several fixed energy points in Fig. 7(d) indicate that the anomaly is markedly different in the energy and momentum space, which is a hallmark feature of correlated electron states. The peaks in the EDCs display prominent dispersionless features at two energy scales marked by arrows. The lowest energy peak persists at all momenta around -80 meV and gradually becomes sharper close to the Γ point. This low-energy feature reveals the formation of the long-lived renormalized quasiparticle. The high-energy feature around −1 eV is considerably broad in both energy and momentum space. In momentum space it attains a large width, demonstrating that these states are significantly incoherent and are created by spectral-weight depletion near −500 meV due to the coupling to spin-fluctuations.

## B. Experimental self-energy

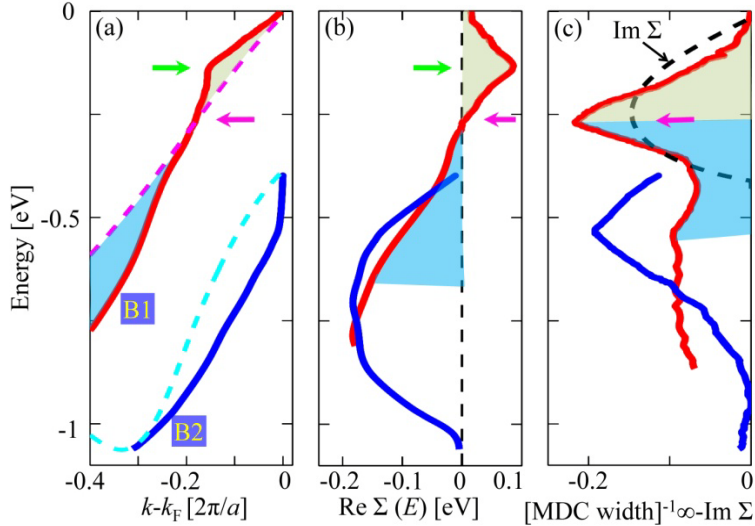

FIG. 8. The real and imaginary parts of the experimental self-energy. (a) Two low-lying dispersions derived from the peak positions of the MDCs [Fig. 7(c)] along M→Γ. The momentum axis is offset by $k_F$ for band B1. Dashed lines of same color give the *ab-initio* dispersion (offset from their corresponding Fermi momenta values), which help expose the degree of anomaly for each dispersion with respect to their linear dispersion features. (b) The real part of the self-energy measures the deviation of each dispersion curve in (a) from its corresponding GGA value. Green arrow marks the location of the low-energy kink. (c) The inverse of the MDCs width obtained with the help of Fig. 7(c), which is proportional to the imaginary part of the self-energy, when the extrinsic background contribution is negligible. Different color fillings in all three curves (corresponding to band B1) separate the characteristic energy scale (marked by magenta arrows). The imaginary part attains a peak at the energy where the real part of the self-energy changes sign. To demonstrate this behavior we calculate Im Σ from Re Σ in (b) using Kramers-Kronig relation and plot is as black dashed line in (c). Reproduced from Ref. 16.

Embedded in the one-electron self-energy is considerable information of the many-body correlation mechanism. In principle, this makes it a very useful quantity to compare with predictions by different theories to identify the appropriate approximation scheme for treating correlation effects. The determination of the self-energy in actinides has so far only been achieved for $USb_2$[62] and $UCoGa_5$.[16] Here we illustrate how to extract two quasiparticle dispersion branches by tracing the locii of the MDC peaks as shown in Fig. 8(a). To study the dispersion renormalization, the common practice is to observe the departure of the dispersion with respect to the nearest LDA one, $\xi_{nk}$ (shown by dashed lines of same color). A visual comparison reveals a sharp kink in the metallic actinide $UCoGa_5$, that is, a change in the slope of the dispersion. The low-energy kink lies around −130 meV (indicated by green arrow), which is higher than that in cuprates (−70 meV).[63] On the other hand, the kink observed in the actinide $USb_2$ is at much lower energy, and thus is most likely caused by electron-phonon coupling.[62] Furthermore, the high-energy kink (−1 eV) poses a ubiquitous anomaly in the 115s similar to the cuprates.[56–58] On the basis of these comparisons, we deduce that the present kink lies well above the phonon energy scale of ∼30 meV for metallic $UCoGa_5$,[51] and hitherto provides a novel system to study the evolution of spin fluctuations in *f*-electron systems.

The experimental values of the real and imaginary parts of the self-energy can be derived from the ARPES spectra by tracing the peak positions of each band $k_n$ as $\Sigma'(k_n,\omega) = \omega - \xi_k^n$, $\Sigma''(k,\omega) \propto 1/\text{FWHM}_{\text{MDC}}$. Here $\xi_k^n$ is $n^{th}$ the GGA dispersion with respect to its corresponding Fermi energy. The obtained results are shown in Figs. 8(b) and 8(c), respectively, for the two extracted bands shown in Fig. 8(a). Taking advantage of the high-precision data, we expose two energy scales in $\Sigma'$ (marked by arrows) which are intrinsically linked to the quasiparticle lifetime $\tau = \hbar/(2\Sigma'')$. In addition to the peak at the kink energy, $\Sigma'$ possesses a sign reversal around 260 meV. This is an important feature, which imposes the constraint that the corresponding $\Sigma''$ should yield a peak exactly at the same energy, due to the Kramers-Kronig relationship. This is indeed in agreement with our findings for UCoGa$_5$ in Fig. 8(c), which exhibits a sharp peak in the extracted broadening, exactly where $\Sigma'$ (changes sign (marked by green arrow). It is worthwhile to mention that in ARPES measurements, the extrinsic source of quasiparticle broadening is typically known to be quasi-linear in energy,[64] while the presence of a peak is a definitive signature of the intrinsic origin of broadening.

## C. Comparison with cuprates and ruthenates

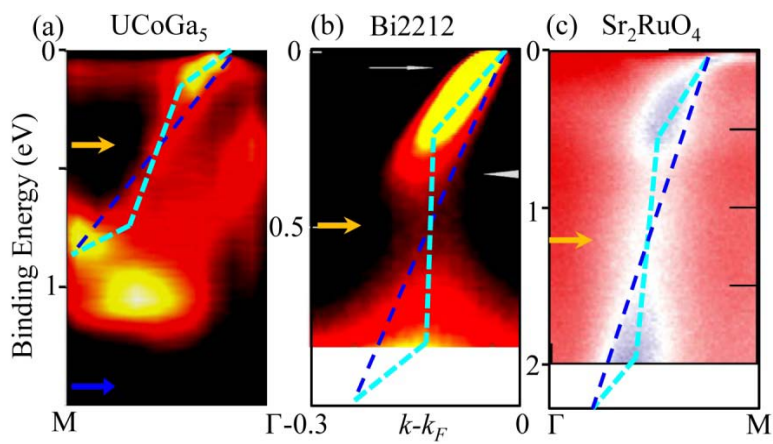

FIG. 9. The ARPES data for various compounds, as different as non-superconducting actinide UCoGa$_5$,[16] high-temperature cuprate superconductor Bi$_2$Sr$_2$CuO$_{8+\delta}$,[56] and odd-parity superconductor Sr$_2$RuO$_4$.[61] The cyan-dashed line guides the eyes on how the experimental dispersion deviates from the linear bare dispersion (blue-dashed line). Arrows indicate various characteristic energy scales (see text).

In Fig. 9, we demonstrate the similarity of the dispersion anomaly in very different systems. The basic concept underlying the lineshape of quasiparticle dispersions is that the mass renormalization is strongly energy dependent and there exists a crossover where the renormalization changes from positive to negative. This part is illustrated by drawing straight lines on top of the ARPES spectral function and studying how they deviate from an idealized linear dispersion connecting the top and bottom of the experimental lineshape. We find a surprisingly similar behavior in all three systems, and following the self-energy analysis illustrated in Fig. 8, we can easily see that the extracted real and imaginary parts of the self-energy have similar characteristics as obtained in Fig. 8(b) and Fig. 8(c). This means, $\Sigma' = 0$ where $\Sigma''$ peaks, and the dispersion below and above this energy scale are renormalized towards and away from the Fermi level, respectively. Without loss of generality, we attribute the origin of this peak in $\Sigma''$ to be due to strong fluctuations (most likely due to spin fluctuations as in actinides[6] and cuprates[49,59]), although the microscopic origin of fluctuations can be different. These two energy scales yield a ubiquitous peak-dip-hump feature, which is present in a number of actinide systems.

### D. Photoemission spectroscopy

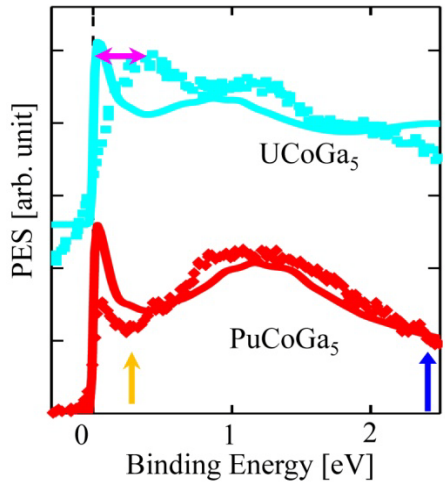

FIG. 10. Computed angle-integrated photoemission spectra (solid lines) for two actinide materials compared with corresponding experimental data (symbols of same color). The results for UCoGa$_5$ are shifted vertically upward by arbitrary value for visualization. The relative intensity difference between peak and hump depends on various experimental conditions such as photon energy and background. The experimental data of UCoGa$_5$ is obtained via x-ray photoemission spectroscopy (XPS),[65] whereas that for PuCoGa$_5$ is from synchrotron data.[17]

The angle-integrated photoemission spectroscopy can directly probe the DOS of the filled states below the Fermi level to the extent that matrix-element effects are not important.[59] To validate our results, we directly compare them with available data for PuCoGa$_5$[17] at 77 K, and x-ray photoemission data of UCoGa$_5$[65] in Fig. 10. The theory data are reproduced from Fig. 6. Our computed results are in good agreement with the PES experiments, revealing a ubiquitous presence of the peak-dip-hump feature or equivalently of electronic hot spots in the momentum-energy space of actinides. Near $E_F$ the experiment shows a broader feature than theory with less spectral weight, which may be related to experimental resolution and theoretical approximations. The present calculation slightly underestimates the dip in the spectral weight, which stems from the neglect of orbital matrix-elements, charge and other fluctuations, as well as the quasiparticle approximation in the self-consistency scheme of the calculation of the self-energy. The key result is that both the spectral weight loss at low energy and high energy are well captured by the spin-fluctuation model.

The PES data for other actinide based compounds, such as α and δ-phase stabilized Pu,[66–69] UNiGa$_5$[70] also exhibit similar peak-dip-hump features. These results suggest that intermetallic actinide compounds and elemental Pu reside in the intermediate Coulomb interaction regime.[6] In addition, it should be noted that the intermediate Coulomb interaction scenario also successfully explains many salient features of high-temperature copper-oxide superconductors,[5,71] where spin fluctuations are thought to be important.

### E. Fermiology and de Haas-van Alphen oscillations

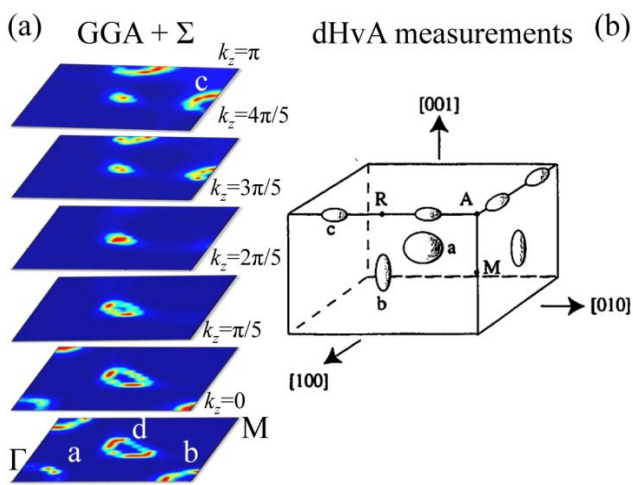

FIG. 11. FS maps and dHvA results of UCoGa$_5$. (a) FS cuts at six representative $k_z$-values after including the self-energy ($\Sigma$) correction. (b) Measured FS pockets from dHvA oscillations.[72] In the GGA+$\Sigma$ and dHvA data all the FS pieces labeled 'a', 'b', and 'c' are well reproduced, while the large tube-like FS 'd' obtained in our calculation around ($\pi/2$, $\pi/2$, 0) is not seen in the dHvA experiment. Reproduced from Ref. 16.

The fermiology of the self-energy dressed FS topology is presented for several $k_z$ cuts in Fig. 11(a). We note that the self-energy corrections in the GGA+$\Sigma$ calculation of the dispersion do not make a dramatic change to the overall GGA FS maps (shown in Fig. 2). The GGA computed FS maps are in accord with de-Haas-van Alphen (dHvA) measurements, except for the tube-like FS around ($\pi/2$, $\pi/2$, 0).[72] In the $k_z = 0$ plane, we obtain three FSs: at the Γ point (labeled 'a' FS), at ($\pi$, 0, 0) (labeled 'b' FS), and around ($\pi/2$, $\pi/2$, 0) connecting the 'Σ' to 'S' points (labeled 'd' FS), and their equivalent points in the Brillouin zone. In the $k_z = \pi$ plane, the 'a' and 'b' FSs disappear, whereas a fourth FS 'c' reaches maximum size. After including the spin-fluctuation interaction in the GGA+$\Sigma$ calculation, the spectral weight across each FS varies, but the general shape and position of each FS remains very much the same, satisfying the Luttinger theorem for conserved FS volume. Only the tube-shaped 'd' pocket changed from closed to open orbits along the $k_z$ direction due to the small shift of the chemical potential to lower energies. This sensitivity to small changes in the electronic structure calculation is consistent with earlier reports of closed versus open tube-like pockets.[54,55] The 'a', 'b' and 'c' pockets, seen in the dHvA measurements, are in qualitative agreement with electronic structure calculations, as previously discussed by Opahle et al.[54], while the 'd' FS was not observed. It is possible that the open-loop pocket

'd' was masked by the relatively broad fundamental frequency of the 'a' orbit. A targeted search for the associated dHvA oscillation should resolve the so far missing tube-shaped pocket.

## V. DISCUSSION

It is important to note that all energy scales of the renormalization and corresponding spectral weight of electronic dispersion are related to each other. In other words, the spin-fluctuation induced anomaly at energies as high as 500 meV, is also responsible for mass renormalization at low energy, and thus in turn affects superconductivity. To see that connection, we estimate the coupling constant (mass renormalization) and compare with experiments.

### A. Spin-fluctuation coupling constant and implication to superconductivity

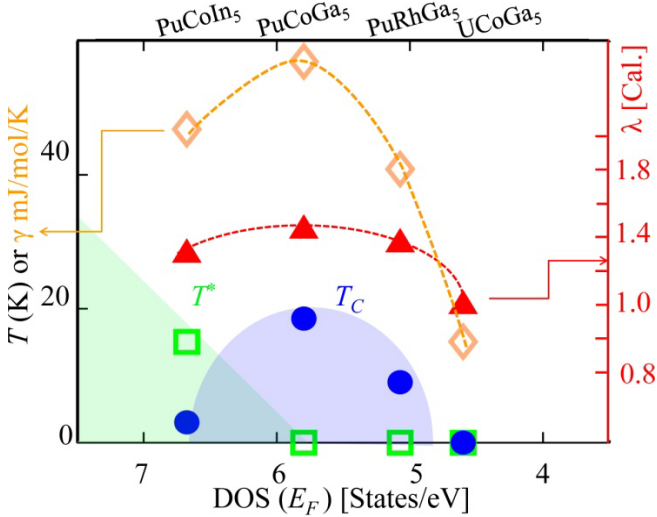

FIG. 12. Experimental values of $T_c$ and the antiferromagnetic transition $T^*$ as a function of the GGA DOS at $E_F$ (theory) and compared with computed values of the spin-fluctuation coupling constant $\lambda$ and corresponding Sommerfeld coefficient $\gamma$. Reproduced from Ref. 6.

We calculate the spin-fluctuation coupling constant $\lambda$ from the energy derivative of $\Sigma'$. In the low-energy region, we obtain $\Sigma'(\xi_k) \approx -\lambda \xi_k = (1 - Z^{-1})\xi_k$. The coupling constant $\lambda$ follows the same material dependence as $T_c$ across the series from PuCoIn$_5 \rightarrow$ UCoGa$_5$ with its maximum for PuCoGa$_5$. Although $\lambda$ is quite large for PuCoIn$_5$, its $T_c$ is strongly suppressed probably due to competition with a significantly large impurity phase of PuIn$_3$. In addition, Bauer et al.[13] reported a three to four times larger spin-fluctuation energy scale between PuCoIn$_5$ and PuCoGa$_5$ by extracting a spin-fluctuation temperature from the Sommerfeld coefficient or maximum in resistivity. They further concluded that PuCoIn$_5$ exhibits more localized 5f-electron physics due to the 28% enlarged unit cell.[13] In contrast, our self-consistent self-energy calculations do not provide any evidence for these interpretations. We do not see noticeably different spin-fluctuation coupling constants between both materials, see Fig. 12. Therefore the origin of these discrepancies may be tied to the magnetic properties of the impurity phase PuIn$_3$.

Encouraged by the aforementioned analysis, we now focus on delineating the nature of the elementary excitations responsible for this anomaly. As mentioned before, both the energy scales and the strength of electron-phonon interaction[54,55] is insufficient to capture our observed dispersion anomalies. We demonstrated that the electron-electron interaction due to dynamical spin fluctuations of exchanged bosons can describe the data. In UCoGa$_5$ strong spin-orbit coupling from relativistic effects enables substantial spin fluctuations whose feedback results in an increase of the electron mass and a shortening of the quasiparticle lifetime.

## B. Comparison with measured Sommerfeld coefficient

### 1. PuCoGa$_5$

Next, we compare our estimation of λ in PuCoGa$_5$ with experimental values. Experimentally, an estimate of λ at the Fermi level is obtained by comparing the measured Sommerfeld coefficient with respect to its non-interacting value, $\gamma \approx (1 + \lambda)\gamma_0$. For PuCoGa$_5$, we find a GGA value of $\gamma_0$=20.5 mJ/mol K$^2$, which leads to the renormalized theoretical value of γ = 54 mJ/mol/K$^2$ with λ =1.4. This calculated Sommerfeld coefficient is only slightly less than the corresponding experimental value of 77 mJ/mol/K$^2$,[11] suggesting that there is room for additional phonon fluctuations of about $\lambda_{ep} \sim 0.8$; a value which is very close to the electron-phonon coupling constant deduced by first-principles calculations ($\lambda_{ep}$ = 0.7).[51,73] Note that our calculated coupling constant of λ =1.4 for PuCoGa$_5$ is smaller than the calculated value of 2.5 obtained within the LDA+DMFT approximation.[42]

### 2. UCoGa$_5$

Following the same procedure, we obtain for UCoGa$_5$ the theoretical estimate of the spin-fluctuation coupling constant λ =1. The experimental values of γ are between γ =10-21 mJ/mol K$^2$,[8,74] while the GGA calculated value is $\gamma_0$=5.8 mJ/mol K$^2$. The theoretical Sommerfeld coefficient agrees well with the literature[54,55] and thus gives an experimental estimate of $\lambda \approx 0.7 - 2.6$. Therefore, the combination of the calculated spin-fluctuation coupling[51,73] constant of λ =1.0 with the low-energy electron-phonon coupling constant, same $\lambda_{ep}$~0.7 as for PuCoGa$_5$, adequately describes the observed modest electronic mass renormalization in UCoGa$_5$.

# VI. SUMMARY AND CONCLUSION

Among the several computational advantages that the intermediate Coulomb-$U$ coupling method provides for modeling of real materials is the $GW$-like perturbative approach with spin fluctuations to actinides, because it has its foundation on a diagrammatic description of fluctuations. In this work we focused on spin fluctuations, but the formulation is more general to account for charge and pair fluctuations as well. An important outcome of this study is the agreement of the self-consistently calculated results with various photoemission, de Haas-van Alphen, and heat capacity measurements. We conclude with three key observations: (1) The perturbative approach requires that the interaction part or the value of the Coulomb potential $U$ is smaller than the non-interacting electronic bandwidth $D$. Since we start with the first-principles band structure, we can choose a value of $U$ up to a value below the GGA bandwidth $D$ as long as it does not introduce an instability in the RPA susceptibility to an ordered ground state. In the intermediate Coulomb interaction scheme, we fix $U_i \sim D_i$ for each band (orbital) $i$. (2) The mass renormalization of the quasiparticle obtained via the self-energy $\Sigma$ at the Fermi level is related to the shape of $\Sigma''$, that is, to the quasiparticle lifetime due to the Kramers-Kronig relationship. The energy scale and the strength of $\Sigma''$ are the key factors that determine the characteristics of the correlation effects of a given material. In the weak-coupling regime, the peak height of $\Sigma''$ is typically less pronounced and thus often gives a negligible amount to the width of the quasiparticle lineshape. As this study has shown, the quasiparticle states near the Fermi level in the 115 actinides can be well described within the itinerant picture with a modest mass renormalization, consistent with Fermi liquid theory and spin fluctuations. (3) In strong-coupling approaches like the DMFT method, the peak in $\Sigma''$ is pronounced, thus it may lead to an overestimation of the spectral weight loss in the low-energy region. However, more importantly single-site DMFT lacks the momentum dependence of the self-energy compared to the $GW$ method.

In summary, the intermediate Coulomb-$U$ coupling method offers an opportunity to address the complex nature of the dynamic band renormalization in actinides. This is not unexpected for materials with itinerant electrons, since similar theories based on the coupling between spin fluctuations and electrons are among the leading contenders for explaining the origin of high-temperature superconductivity,[75] itinerant magnetism, and other emergent states of matter in $d$- and $f$-electron systems.[7,76] In addition, spin fluctuations in 5$f$-electron systems are a manifestation of relativistic effects due to spin-orbit split states of order ~0.5 eV. In fact, spin-orbit coupling may provide a novel electronic knob to tune the interaction strength and characteristic frequency of the mediating boson. We hope that our results for the self-energy will motivate efforts to identify generic dispersion anomalies and 'waterfall' physics in a larger class of actinide and heavy-fermion compounds.

In the future it will be interesting to extend the intermediate coupling calculation to a wider range of the 115 family[14,15,77] and the so-called 122 family[78,79] of actinides to understand and predict their correlated electronic structure and anomalous ground state properties. It should be noted that the unsolved mysteries in the 122 materials like the 'hidden order' phase in $URu_2Si_2$[38,80], or the unusual Fermi surface reconstruction in $YbRh_2Si_2$[81], call for models that are sophisticated enough to incorporate the high-energy fluctuation phenomena and strong spin-orbit coupling for generating spectral weight 'hot-spots', which are responsible for the emergence of broken symmetry phases[80] in the low-energy region.

## ACKNOWLEDGMENTS

We thank T. Durakiewicz, J. J. Joyce, A.V. Balatsky, R. S. Markiewicz, A. Bansil, P. Werner, P. Oppeneer, F. Ronning, and E.D. Bauer for discussions. Work at the Los Alamos National Laboratory was supported by the U.S. DOE under Contract No. DE-AC52-06NA25396 through the LDRD Program. We acknowledge computing allocations by NERSC through the Office of Science (BES) under Contract No. DE-AC02-05CH11231.

## REFERENCES

[1] P. W. Anderson, *Localized magnetic states in metals.* Phys. Rev. **124**, 41 (1961).

[2] A. C. Hewson, *The Kondo problem to heavy fermions.* (Cambridge Univ. Press) (1983).

[3] Q. Si, E. Abrahams, J. Dai, and J.-X. Zhu, *Correlation effect in the iron pnictides*. New J. Phys. **11**, 045001 (2009).

[4] N. J. Curro, T. Caldwell, E. D. Bauer, L. A. Morales, M. J. Graf, Y. Bang, A. V. Balatsky, J. D. Thompson, and J. L. Sarrao, *Unconventional superconductivity in $PuCoGa_5$.* Nature **434**, 622 (2005).

[5] T. Das, R. S. Markiewicz, and A. Bansil, *Optical model-solution to the competition between a pseudogap phase and a charge-transfer-gap phase in high-temperature cuprate superconductors.* Phys. Rev. B **81**, 174504 (2010).

[6] T. Das, J.-X. Zhu, and M. J. Graf, *Spin fluctuations and the peak-dip-hump feature in the photoemission spectrum of actinides.* Phys. Rev. Lett. **108**, 017001 (2012).

[7] T. Moriya, *Spin fluctuations in itinerant electron magnetism* (Springer, Berlin) (1985).

[8] N. O. Moreno, E. D. Bauer, J. L. Sarrao, M. F. Hundley, J. D. Thompson, and Z. Fisk, *Thermodynamic and*

transport properties of single-crystalline UMGa$_5$ (M=Fe, Co, Ni, Ru, Rh, Pd, Os, Ir, Pt)*, Phys. Rev. B **72**, 035119 (2005).

[9] K. Kanekoa, N. Metokia, G. H. Landera, N. Bernhoeftd, Y. Tokiwaa, Y. Hagaa, Y. Onukia, Y. Ishiia, *Neutron diffraction study of 5f itinerant antiferromagnet UPtGa$_5$ and UNiGa$_5$*, Physica B: Condensed Matter **329-333**, 510-511 (2003).

[10] P. Boulet, E. Colineau, F. Wastin, J. Rebizant, P. Javorsk´y, G. H. Lander, and J. D. Thompson, *Tuning of the electronic properties in PuCoGa$_5$ by actinide (U, Np) and transition-metal (Fe, Rh, Ni) substitutions*. Phys. Rev. B **72**, 104508 (2005).

[11] J. L. Sarrao, L. A. Morales, J. D. Thompson, B. L. Scott, G. R. Stewart, F. Wastin, J. Rebizant, P. Boulet, E. Colineau and G.H. Lander, *Plutonium-based superconductivity with a transition temperature above 18 K*. Nature **420**, 297 (2002).

[12] F. Wastin, P Boulet, J Rebizant, E Colineau and G. H. Lander, *Advances in the preparation and characterization of transuranium systems*. J. Phys.: Condens. Matter **15**, S2279 (2003).

[13] E. D. Bauer, M. M. Altarawneh, P. H. Tobash, K. Gofryk, O. E. Ayala-Valenzuela, J. N. Mitchell, R. D. McDonald, C. H. Mielke, F. Ronning, J.-C. Griveau, E. Colineau, R. Eloirdi, R. Caciuffo, B. L. Scott, O. Janka, S. M. Kauzlarich, and J. D. Thompson, *Localized 5f electrons in superconducting PuCoIn5: consequences for superconductivity in PuCoGa$_5$*. J. Phys.: Consens. Matter **24**, 052206 (2012).

[14] Y. N. Grin, P. Rogl, and K. Hiebl, *Structural chemistry and magnetic behavior of ternary uranium gallides U(Fe, Co, Ni, Ru, Rh, Pd, Os, Ir, Pt)Ga$_5$*, J. Less. Comm. Met. **121**, 497 (1986).

[15] S. Ikeda, Y. Tokiwa,T. Okubo, M. Yamada, T. D. Matsuda, Y. Inada, R. Settai, E. Yamamoto, Y. Haga, and Y. Onuki, *Magnetic and Fermi surface properties of UTGa$_5$ (T : Fe, Co and Pt)*, Physica B **329-333**, 610 (2003).

[16] T. Das, T. Durakiewicz, J.-X. Zhu, J. J. Joyce, J. L. Sarrao, and M. J. Graf, *Imaging the formation of high-energy dispersion anomaly in the actinide UCoGa$_5$*. arXiv:1206.1302. Accepted for publication in PRX (2012).

[17] J. J. Joyce, J. M.Wills, T. Durakiewicz, M. T. Butterfield, E. Guziewicz, J. L. Sarrao, L. A. Morales, A. J. Arko, and O. Eriksson, *Photoemission and the electronic structure of PuCoGa$_5$,* Phys. Rev. Lett. **91**, 176401 (2003); J. J. Joyce, T. Durakiewicz, K. S. Graham, E. D. Bauer, D. P Moore, J. N. Mitchell, J. A. Kennison, R. L. Martin, L. E. Roy, and G. E. Scuseria, *Pu electronic structure and photoemission spectroscopy*. J. Phys.: Conf. Series **273**,

012023 (2011).

[18] S. Fujimori, K. Terai, Y. Takeda, T. Okane, Y. Saitoh, Y. Muramatsu, A. Fujimori, H. Yamagami, Y. Tokiwa, S. Ikeda, T. D. Matsuda, Y. Haga, E. Yamamoto, and Y. Onuki, *Itinerant U 5f band states in the layered compound UFeGa$_5$ observed by soft x-ray angle-resolved photoemission spectroscopy.* Phys. Rev. B **73**, 125109 (2006).

[19] M. E. Pezzoli, K. Haule, and G. Kotliar, *Neutron magnetic form factor in strongly correlated materials*, Phys. Rev. Lett. **106**, 016403 (2011).

[20] J.-X. Zhu, P. H. Tobash, E. D. Bauer, F. Ronning, B. L. Scott, K. Haule, G. Kotliar, R. C. Albers, and J. M. Wills, *Electronic structure and correlation effects in PuCoIn$_5$ as compared to PuCoGa$_5$.* Europhys. Lett. **97**, 57001 (2012).

[21] R. M. Martin, *Electronic structure: Basic theory and practical methods.* (Cambridge University Press, Cambridge, (2004).

[22] P. Hohenberg, and W. Kohn, *Inhomogeneous electron gas*, Phys. Rev. **136**, B864 (1964).

[23] W. Kohn, and L. J. Sham, *Self-consistent equations including exchange and correlation effects*, Phys. Rev. **140**, A1133 (1965).

[24] J. P. Perdew, K. Burke, and M. Ernzerhof, *Generalized gradient approximation made simple*, Phys. Rev. Lett. **77**, 3865 (1996); Erratum: **78**, 1396 (1997).

[25] V. I. Anisimov, F. Aryasetiawan, and A. I. Lichtenstein, *First-principles calculations of the electronic structure and spectra of strongly correlated systems: dynamical mean-field theory.* J. Phys.: Condens. Matter **9**, 767 (1997).

[26] J. P. Perdew, M. Ernzerhof, and K. Burke, *Rationale for mixing exact exchange with density functional approximations.* J. Chem. Phys. **105**, 9982 (1996).

[27] J. P. Perdew, and A. Zunger, *Self-interaction correction to density-functional approximations for many-electron systems*, Phys. Rev. B **23**, 5048 (1981).

[28] A. Svane, and O. Gunnarsson, *Localization in the self-interaction-corrected density-functional formalism,* Phys. Rev. B **37**, 9919 (1988).

[29] Z. Szotek, W. M. Temmerman, and H. Winter, *Application of the self-interaction correction to transition-metal oxides*. Phys. Rev. B **47**, 4029 (1993).

[30] F. Aryasetiawan and O. Gunnarsson, *The GW method,* Rep. Prog. Phys. **61**, 237 (1998).

[31] L. Hedin, *New method for calculating the one-particle Green's function with application to the electron-gas problem*, Phys. Rev. **139**, A796 (1965).

[32] W. Metzner and D. Vollhardt, *Correlated lattice fermions in $d=\infty$ dimensions*, Phys. Rev. Lett. **62**, 324 (1989); Erratum: **62**, 1066 (1989).

[33] E. Müller-Hartmann, *The Hubbard model at high dimensions: some exact results and weak coupling theory*, Z. Phys. B-Condens. Matter **76**, 211 (1989).

[34] A. Georges, G. Kotliar, W. Krauth, and M. Rozenberg, *Dynamical mean-field theory of strongly correlated fermion systems and the limit of infinite dimensions*, Rev. Mod. Phys. **68**, 13 (1996).

[35] G. Kotliar, S. Y. Savrasov, K. Haule, V. S. Oudovenko, O. Parcollet, and C. A. Marianetti, *Electronic structure calculations with dynamical mean-field theory*, Rev. Mod. Phys. **78**, 865 (2006).

[36] G. Onida, L. Reining, and A. Rubio, *Electronic excitations: density-functional versus many-body Green's function approaches*, Rev. Mod. Phys. **74**, 601 (2002).

[37] K. Held, *Electronic structure calculations using dynamical mean field theory*, Adv. Phys. **56**, 829 (2007).

[38] K. Haule and G. Kotliar, *Arrested Kondo effect and hidden order in $URu_2Si_2$*, Nat. Phys. **5**, 796 (2009).

[39] J. H. Shim, K. Haule, and G. Kotliar, *Modelling the localized to itinerant electronic transition in the heavy fermion system $CeIrIn_5$*, Science **318**, 1615 (2007).

[40] H. C. Choi, B. I. Min, J. H. Shim, K. Haule, and G. Kotliar, *Temperature-dependent Fermi surface evolution in heavy fermion $CeIrIn_5$*, Phys. Rev. Lett. **108**, 016402 (2012).

[41] L. V. Pourovskii, M. I. Katsnelson, and A. I. Lichtenstein, *Correlation effects in electronic structure of $PuCoGa_5$,* Phys. Rev. B **73**, 060506(R) (2006).

[42] A. B. Shick, Ján Rusz, J. Kolorenc, P. M. Oppeneer, and L. Havela, *Theoretical investigation of electronic structure, electric field gradients, and photoemission of $PuCoGa_5$ and $PuRhGa_5$ superconductors*, Phys. Rev. B **83**, 155105 (2011).

[43] J. P. Perdew, S. Burke, and M. Ernzerhof, *Generalized gradient approximation made simple*. Phys. Rev. Lett. **77**, 3865 (1996).

[44] T. Takimoto, T. Hotta, and K. Ueda, *Strong-coupling theory of superconductivity in a degenerate Hubbard model.*

Phys. Rev. B **69**, 104504 (2004).

[45] T. Takimoto, T. Hotta, K. Ueda, *Strong-coupling theory of superconductivity in a degenerate Hubbard model,* Phys. Rev. B **69**, 104504 (2004).

[46] T. Das and A. V. Balatsky, *Two energy scales in the magnetic resonance spectrum of electron and hole doped pnictide superconductors,* Phys. Rev. Lett. **106**, 157004 (2011).

[47] T. Das and A. V. Balatsky, *Stripes, spin resonance, and nodeless d-wave pairing symmetry in $Fe_2Se_2$-based layered superconductors.* Phys. Rev. B **84**, 014521 (2011); T. Das and A. V. Balatsky, *Modulated superconductivity due to vacancy and magnetic order in $A_yFe_{2x-2}Se_2$ [A=Cs, K, (Tl,Rb), (Tl,K)] iron-selenide superconductors.* Phys. Rev. B **84**, 115117 (2011).

[48] N. E. Bickers, D. J. Scalapino, and S. R. White, *Conserving approximations for strongly correlated electron systems: Bethe-Salpeter equation and dynamics for the two-dimensional Hubbard model*, Phys. Rev. Lett. **62**, 961 (1989).

[49] R. S. Markiewicz, S. Sahrakorpi, and A. Bansil, *Paramagnon-induced dispersion anomalies in the cuprates*. Phys. Rev. B **76**, 174514 (2007).

[50] J. C. Ward, *An identity in quantum electrodynamics*, Phys. Rev. **78**, 182 (1950).

[51] P. Piekarz, K. Parlinski, P. T. Jochym, A. M. Olés, J.-P. Sanchez, and J. Rebizant, *First-principles study of phonon modes in $PuCoGa_5$ superconductor.* Phys. Rev. B **72**, 014521 (2005).

[52] P. Blaha, K. Schwarz, G.K.H. Madsen, D. Kvasnicka, and J. Luitz, *An augmented plane wave + local orbitals program for calculating crystal properties,* (K. Schwarz, Tech. Universit¨at Wien, Austria, 2001).

[53] J. Kunes, P. Novák, R. Schmid, P. Blaha, and K. Schwarz, *Electronic structure of fcc Th: Spin-orbit calculation with $6p_{1/2}$ local orbital extension*, Phys. Rev. B **64**, 153102 (2001).

[54] I. Ophale, S. Elgazzar, K. Koepernik, and P. M. Oppeneer, *Electronic structure of the Pu-based superconductor $PuCoGa_5$ and of related actinide-115 compounds.* Phys. Rev. B **70**, 104504 (2004).

[55] T. Maehira, T. Hotta, K. Ueda, and A. Hasegawa, *Electronic structure and the Fermi surface of $PuCoGa_5$ and $NpCoGa_5$*, Phys. Rev. Lett. **90**, 207007 (2003).

[56] J. Graf, G.-H. Gweon, K. McElroy, S. Y. Zhou, C. Jozwiak, E. Rotenberg, A. Bill, T. Sasagawa, H. Eisaki, S. Uchida, H. Takagi, D.-H. Lee, and A. Lanzara, *Universal high energy anomaly in the angle-resolved*

*photoemission spectra of high temperature superconductors: possible evidence of spinon and holon branches.* Phys. Rev. Lett. **98**, 067004 (2007).

[57] B. P. Xie, K. Yang, D. W. Shen, J. F. Zhao, H. W. Ou, J. Wei, S. Y. Gu, M. Arita, S. Qiao, H. Namatame, M. Taniguchi, N. Kaneko, H. Eisaki, K. D. Tsuei, C. M. Cheng, I. Vobornik, J. Fujii, G. Rossi, Z. Q. Yang, and D. L. Feng, *High-energy scale revival and giant kink in the dispersion of a cuprate superconductor*, Phys. Rev. Lett. **98**, 147001 (2007).

[58] T. Valla, T. E. Kidd, W.-G. Yin, G. D. Gu, P. D. Johnson, Z.-H. Pan, and A. V. Fedorov, *High-energy kink observed in the electron dispersion of high-temperature cuprate superconductors*, Phys. Rev. Lett. **98**, 167003 (2007).

[59] S. Basak, T. Das, H. Lin, J. Nieminen, M. Lindroos, R. S. Markiewicz, and A. Bansil, *Origin of the high-energy kink in the photoemission spectrum of the high-temperature superconductor $Bi_2Sr_2CaCu_2O_8$.* Phys. Rev. B **80**, 214520 (2009).

[60] J. J. Yeh and I. Lindau, At. Data Nucl. Data Tables **32**, 1 (1985).

[61] H. Iwasawa, Y. Yoshida, I. Hase, K. Shimada, H. Namatame, M. Taniguchi, and Y. Aiura, *High-Energy Anomaly in the Band Dispersion of the Ruthenate Superconductor.* Phys. Rev. Lett. **109**, 066404 (2012).

[62] T. Durakiewicz, P. S. Riseborough, C. G. Olson, J. J. Joyce, P. M. Oppeneer, S. Elgazzar, E. D. Bauer, J. L. Sarrao, E. Guziewicz, D. P. Moore, M. T. Butterfield, and K. S. Graham, *Observation of a kink in the dispersion of f-electrons.* Europhys. Lett. **84**, 37003 (2008).

[63] A. Lanzara, P. V. Bogdanov, X. J. Zhou, S. A. Kellar, D. L. Feng, E. D. Lu, T. Yoshida, H. Eisaki, A. Fujimori K. Kishio, J.-I. Shimoyama, T. Nodak, S. Uchidak, Z. Hussain, and Z.-X. Shen, *Evidence for ubiquitous strong electron-phonon coupling in high-temperature superconductors.* Nature **412**, 510 (2001).

[64] D. A. Shirley, *High-resolution x-ray photoemission spectrum of the valence bands of gold.* Phys. Rev. B **5**, 4709 (1972).

[65] R. Troc, Z. Bukowski, C. Sukowski, H. Misiorek, J. A. Morkowski and A. Szajek, G. Chekowska, *Electronic structure, magnetic, and transport studies of single-crystalline $UCoGa_5$.* Phys. Rev. B **70**, 184443 (2004).

[66] J.M. Wills, O. Eriksson, A. Delin, P. H Andersson, J. J. Joyce, T Durakiewicz, M.T Butterfield, A.J Arko, D.P Moore, and L.A Morales, *A novel electronic cofiguration of the 5f states in δ -plutonium as revealed by the*

*photo-electron spectra.* J. Elec. Spec. Rel. Phenom. **135**, 163-166 (2004).

[67] A.J. Arko, J. J. Joyce, L. Morales, J. Wills, J. Lashley, F. Wastin and J. Rebizant, *Electronic structure of α-and δ-Pu from photoelectron spectroscopy.* Phys. Rev. B **62**, 1773 (2000).

[68] S. Y. Savrasov, G. Kotliar, and E. Abrahams, *Correlated electrons in δ-plutonium within a dynamical mean-field picture.* Nature **410**, 793 (2001).

[69] J. H. Shim, K. Haule, and G. Kotliar, *Fluctuating valence in a correlated solid and the anomalous properties of δ-plutonium.* Nature **446**, 513 (2007).

[70] T. Durakiewicz, and J. J. Joyce (private communication).

[71] T. Das, R. S. Markiewicz, and A. Bansil, *Emergence of non-Fermi-liquid behavior due to Fermi surface reconstruction in the underdoped cuprate superconductors.* Phys. Rev. B **81**, 184515 (2010).

[72] S. Ikeda, Y. Tokiwa, T. Okubo, Y. Haga, E. Yamamoto, Y. Inada, R. Settai, and Y. Onuki, *Magnetic and Fermi surface properties of $UCoGa_5$ and $URhGa_5$.* J. Nuc. Sci. Tech. **3**, 206 (2002).

[73] N. Metoki, K. Kaneko, S. Raymond, J-P. Sanchez, P. Piekarz, K. Parlinski, A. M. Oles, S. Ikeda, T. D. Matsuda, Y. Haga, Y. Onuki, and G. H. Lander, *Phonons in $UCoGa_5$*, Physica B **373**, 1003 (2006).

[74] S. Noguchi and K. Okuda *Magnetism of ternary compounds U-T -Ga (T = transition elements)*, J. Magn. Magn. Mat. **104-107**, 57 (1992).

[75] D. J. Scalapino, E. Loh, and J. E. Hirsch, *d-wave pairing near a spin-density-wave instability*, Phys. Rev. B **34**, 8190 (1986).

[76] T. Moriya and T. Takimoto, *Anomalous properties around magnetic instability in heavy-electron systems*, J. Phys. Soc. Jpn. **64**, 960 (1995).

[77] K. Moore and G. van der Laan, *Nature of the 5f states in actinide metals.* Rev. Mod. Phy. **81**, 235 (2009).

[78] T. T. M. Palstra, A. A. Menovsky, J. van den Berg, A. J. Dirkmaat, P. H. Kes, G. J. Nieuwenhuys, and J. A. Mydosh, *Superconducting and magnetic transitions in the heavy-fermion system $URu_2Si_2$.* Phys. Rev. Lett. **55**, 2727 (1985).

[79] O. Trovarelli, C. Geibel, S. Mederle, C. Langhammer, F. M. Grosche, P. Gegenwart, M. Lang, G. Sparn, and F. Steglich, *$YbRh_2Si_2$: Pronounced non-Fermi-liquid effects above a low-lying magnetic phase transition.* Phys. Rev. Lett. **85**, 626 (2000).

[80] T. Das, *Spin-orbit density wave induced hidden topological order in $URu_2Si_2$.* Sci. Rep. **2**, 596 (2012).

[81] S. Paschen, T. Lühmann, S. Wirth, P. Gegenwart, O. Trovarelli, C. Geibel, F. Steglich, P. Coleman, and Q. Si, *Hall-effect evolution across a heavy-fermion quantum critical point.* Nature **432**, 881885 (2004).